\documentclass[aps,prb,twocolumn,superscriptaddress,longbibliography]{revtex4-2}

\usepackage{graphicx}
\usepackage{amsmath,amssymb,bm}
\usepackage[version=4]{mhchem}
\usepackage{booktabs}
\usepackage{siunitx}
\usepackage[T1]{fontenc}
\usepackage[usenames,dvipsnames]{xcolor}
\usepackage[colorlinks=true,linkcolor=BlueViolet,citecolor=ForestGreen,urlcolor=MidnightBlue]{hyperref}
\usepackage{tikz}
\usetikzlibrary{arrows.meta,positioning,calc,decorations.pathreplacing}

\providecommand{\sech}{\operatorname{sech}}

\begin{document}

\title{Dielectric signatures of crystal-field and low-temperature correlated
dynamics in \ce{NdMgAl11O19}}

\author{Sonu Kumar}
\email{sonu.kumar@matfyz.cuni.cz}
\affiliation{Charles University, Faculty of Mathematics and Physics,
Department of Condensed Matter Physics, Prague, Czech Republic}
\affiliation{Adam Mickiewicz University, Faculty of Physics and Astronomy,
Department of Experimental Physics of Condensed Phase, Pozna\'n, Poland}

\author{Ga\"{e}l Bastien}
\affiliation{Charles University, Faculty of Mathematics and Physics,
Department of Condensed Matter Physics, Prague, Czech Republic}

\author{Maxim Savinov}
\affiliation{Institute of Physics, Czech Academy of Sciences, Prague,
Czech Republic}

\author{Ma\l{}gorzata \'Sliwi\'nska-Bartkowiak}
\affiliation{Adam Mickiewicz University, Faculty of Physics and Astronomy,
Department of Experimental Physics of Condensed Phase, Pozna\'n, Poland}

\author{Ross H. Colman}
\affiliation{Charles University, Faculty of Mathematics and Physics,
Department of Condensed Matter Physics, Prague, Czech Republic}

\author{Stanislav Kamba}
\email{kamba@fzu.cz}
\affiliation{Institute of Physics, Czech Academy of Sciences, Prague,
Czech Republic}

\date{\today}

\begin{abstract}
We report dielectric spectroscopy of single-crystalline \ce{NdMgAl11O19}, a
magnetoplumbite hexaaluminate in which localized \ce{Nd^{3+}} moments coexist
with a polarizable \ce{AlO5} bipyramidal network. The real part of the
permittivity, $\varepsilon'_{c}(T)$, measured along the crystallographic $c$
axis, increases as the temperature is lowered from 275~K to 30~K and is
frequency-independent between 4~Hz and 50~kHz. At lower temperatures, a
frequency-dependent decrease in permittivity is observed, followed by a further
upturn below 2~K. The high-frequency $\varepsilon'_{c}(T)$ is described by a
Barrett formula supplemented by an effective two-level contribution, yielding a
robust gap of $\Delta = 25.85 \pm 0.32$~K consistent with the lowest
\ce{Nd^{3+}} crystal-electric-field (CEF) splitting. Below $\sim 30$~K, the
dielectric response becomes strongly frequency and magnetic-field dependent.
Isothermal $\varepsilon_c'(H)$ measurements reveal a reproducible low-field
crossover near $\mu_0H_c \simeq 0.85$~T, which we attribute to the competition
between antiferromagnetic correlations and Zeeman splitting of the ground-state
Kramers doublet. \ce{NdMgAl11O19} thus provides a Kramers reference system in
which dielectric signatures of the excited-state CEF manifold can be
distinguished from those of the field-tuned, correlation-dominated ground-state
doublet sector in a centrosymmetric frustrated magnetoplumbite host.
\end{abstract}

\maketitle

\section{Introduction}

Rare-earth frustrated magnets provide a natural setting in which geometric
frustration, strong spin--orbit coupling, and crystal-electric-field (CEF)
effects combine to produce highly anisotropic low-energy degrees of freedom on
frustrated lattices~\cite{Ramirez2025,Balents2010,Savary2017,Qin2021,Xie2024}.
In such systems, bulk response functions may depend not only on exchange
interactions but also on the structure of the CEF manifold and its coupling to
lattice and orbital degrees of freedom. This broader view is supported by
studies of rare-earth oxides and magnets in which CEF excitations interact with
phonons and other low-energy modes, producing observable anomalies in optical
and magnetodielectric
response~\cite{PhysRevLett.122.177601,PhysRevResearch.2.043169,%
PhysRevResearch.2.023162}. Dielectric measurements can therefore provide
information complementary to conventional magnetic probes when local electronic
and lattice degrees of freedom are significantly coupled.

The magnetoplumbite-derived hexaaluminate family \ce{REMgAl11O19} is
particularly well suited for examining this interplay because frustrated
rare-earth magnetism coexists with a polarizable oxide
framework~\cite{Kumar2026PrMgAl11O19ME,Kumar2025CeMgAl11O19ME,Bastien2024}.
Although the average structure is centrosymmetric, local off-centering within
oxygen bipyramids and the resulting dipolar frustration generate a soft
dielectric background~\cite{Kimura1990_SrM12O19,Iyi1989_PbAl12O19,Li2015,%
wang2014,Li2016_ZNBS,Bastien2024}. The nonmagnetic member \ce{EuAl12O19}
establishes this dipolar host background in the absence of rare-earth CEF
physics~\cite{Bastien2024}. Recent studies on magnetic members of the family
show that the background is not independent of the rare-earth ion: in
\ce{CeMgAl11O19}, a broad field-dependent dielectric anomaly develops on top of
the Barrett-like background and reflects the low-energy Kramers
manifold~\cite{Kumar2025CeMgAl11O19ME}, whereas in \ce{PrMgAl11O19}, local
symmetry lowering and non-Kramers physics produce a qualitatively different
low-temperature dielectric response~\cite{Kumar2026PrMgAl11O19ME}. Taken
together, these results indicate that the dielectric response of this family
provides a sensitive probe of how the single-ion CEF scheme is expressed
through the polar host lattice.

\begin{figure*}[t]
 \centering
 \includegraphics[width=0.98\textwidth]{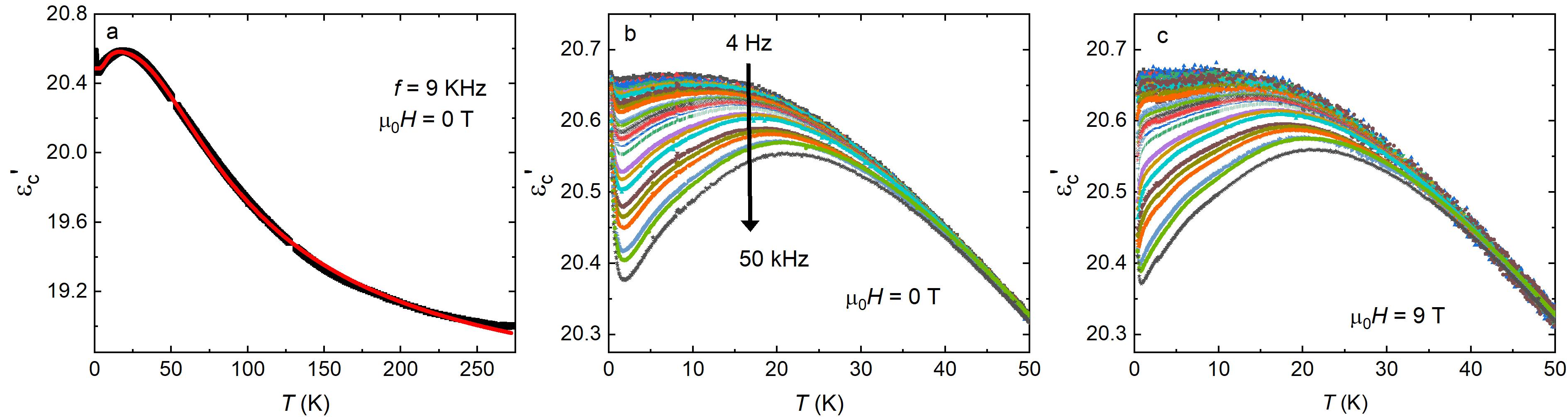}
 \caption{Dielectric response of \ce{NdMgAl11O19}. (a) Zero-field permittivity
 \(\varepsilon_c'(T)\) measured at \(f=9\)~kHz; the solid line is the combined
 fit using a Barrett background and a term associated with thermal depopulation
 of the first excited CEF doublet
 [Eqs.~(\ref{eq:barrett})--(\ref{eq:cef_depop})]. (b,c) Temperature dependence
 of \(\varepsilon_c'(T)\) at selected frequencies in (b) \(\mu_0H=0\) and
 (c) \(\mu_0H=9\)~T.}
 \label{fig:Fig1_eps}
\end{figure*}

\ce{NdMgAl11O19} provides a particularly informative Kramers case within this
series. \ce{Nd^{3+}} is a Kramers ion with \(J=9/2\), and previous magnetic and
thermodynamic work identified \ce{NdMgAl11O19} as a quasi-two-dimensional
triangular-lattice magnet with a strongly anisotropic ground-state doublet,
weak antiferromagnetic interactions (\(\Theta_{\rm CW} \approx -0.5\)~K), and
no conventional long-range order down to the millikelvin
regime~\cite{Kumar2025_NdMgAl}. The same work, supported by point-charge
analysis of the \ce{Nd^{3+}} crystal-field
scheme~\cite{PhysRevResearch.3.023012}, established a low-energy CEF hierarchy
distinct from those of its Ce and Pr analogues, placed the first excited doublet
near \(E_{1}\simeq 2.17\)~meV (\(\approx 25.2\)~K), and identified from
Schottky analysis of the field-evolving low-temperature thermodynamics a
characteristic low-field scale of approximately 0.75--1~T associated with the
ground-state doublet sector. A central question is therefore how these
Nd-specific energy scales appear in the dielectric response.

Here we report dielectric spectroscopy of single-crystalline \ce{NdMgAl11O19}
over broad temperature, frequency, and magnetic-field ranges. We identify three
superposed contributions to the permittivity: a Barrett-like host background, a
CEF-related contribution from the first excited doublet, and a low-temperature
dispersive channel governed by the interplay of antiferromagnetic correlations
and Zeeman splitting of the ground-state doublet.

\section{Experimental Methods}

Single crystals of \ce{NdMgAl11O19} were grown by solid-state reaction followed
by optical floating-zone (OFZ) growth from high-purity \ce{Nd2O3}, \ce{MgO},
and \ce{Al2O3} starting materials. The powders were mixed in stoichiometric
proportions, pressed into feed rods, sintered in air at
1200~$^\circ\mathrm{C}$, and grown in a four-mirror OFZ furnace under flowing
air. Large crystalline grains were mechanically separated and confirmed to be
single crystals by backscattered Laue x-ray diffraction. Further details of
the synthesis and structural characterization are given in
Refs.~\cite{Kumar2025,Kumar2025_NdMgAl,Kumar2026SmMgAl11O19,Bastien2025}.

For dielectric measurements, oriented single-crystalline grains were selected
and cleaved perpendicular to the \(c\) axis. Opposite parallel faces were
polished and coated with thermally evaporated gold electrodes. The complex
dielectric permittivity was measured along the \(c\) axis using an Alpha-AN
impedance analyzer (Novocontrol Technologies) over the frequency range
\SI{1}{\hertz}--\SI{1}{\mega\hertz}. The sample was mounted in a $^3$He
cryostat equipped with a \SI{9}{\tesla} superconducting magnet, with
\(\mathbf{H}\parallel c\). Temperature sweeps were performed between
\SI{0.3}{\kelvin} and \SI{275}{\kelvin} in zero field and between
\SI{0.3}{\kelvin} and \SI{50}{\kelvin} in magnetic fields up to \SI{9}{\tesla}.
At each temperature, the sample was zero-field cooled, thermalized, and the
magnetic field was then set before recording frequency-dependent dielectric
spectra. Further details of the dielectric measurement are given in
Refs.~\cite{Kumar2025CeMgAl11O19ME,Kumar2026PrMgAl11O19ME}.

\section{Results}
\label{sec:results}

Figure~\ref{fig:Fig1_eps}(a) shows the temperature dependence of the real part
of the dielectric permittivity, $\varepsilon_c'(T)$, measured in zero magnetic
field at $f=9$~kHz up to 275~K. On cooling, \(\varepsilon_c'(T)\) increases,
passes through a broad maximum near \(\sim 25\)--30~K, then decreases and
finally shows a further upturn below a few kelvin. Between approximately 35 and
200~K, the response is nearly frequency independent within the experimental
resolution, indicating that this interval is dominated by the intrinsic
high-frequency dielectric background. In this temperature range, the smooth
increase of $\varepsilon_c'(T)$ on cooling is well described by a Barrett
formula [Eq.~(\ref{eq:barrett})], as commonly used in incipient ferroelectrics
and quantum paraelectrics, and may reflect weak softening of a polar phonon.

\begin{equation}
\varepsilon'_{\rm Barrett}(T)=\varepsilon_{\infty}+
\frac{C}{\left(\frac{T_1}{2}\right)\coth\!\left(\frac{T_1}{2T}\right)-T_0},
\label{eq:barrett}
\end{equation}
where \(\varepsilon_{\infty}\) is a temperature-independent background, \(T_1\)
is the characteristic quantum temperature, \(T_0\) is a Curie--Weiss-like
temperature, and the constant \(C\) sets the overall scale of the dielectric
response.

\begin{figure*}[t]
  \centering
  \includegraphics[width=0.9\textwidth]{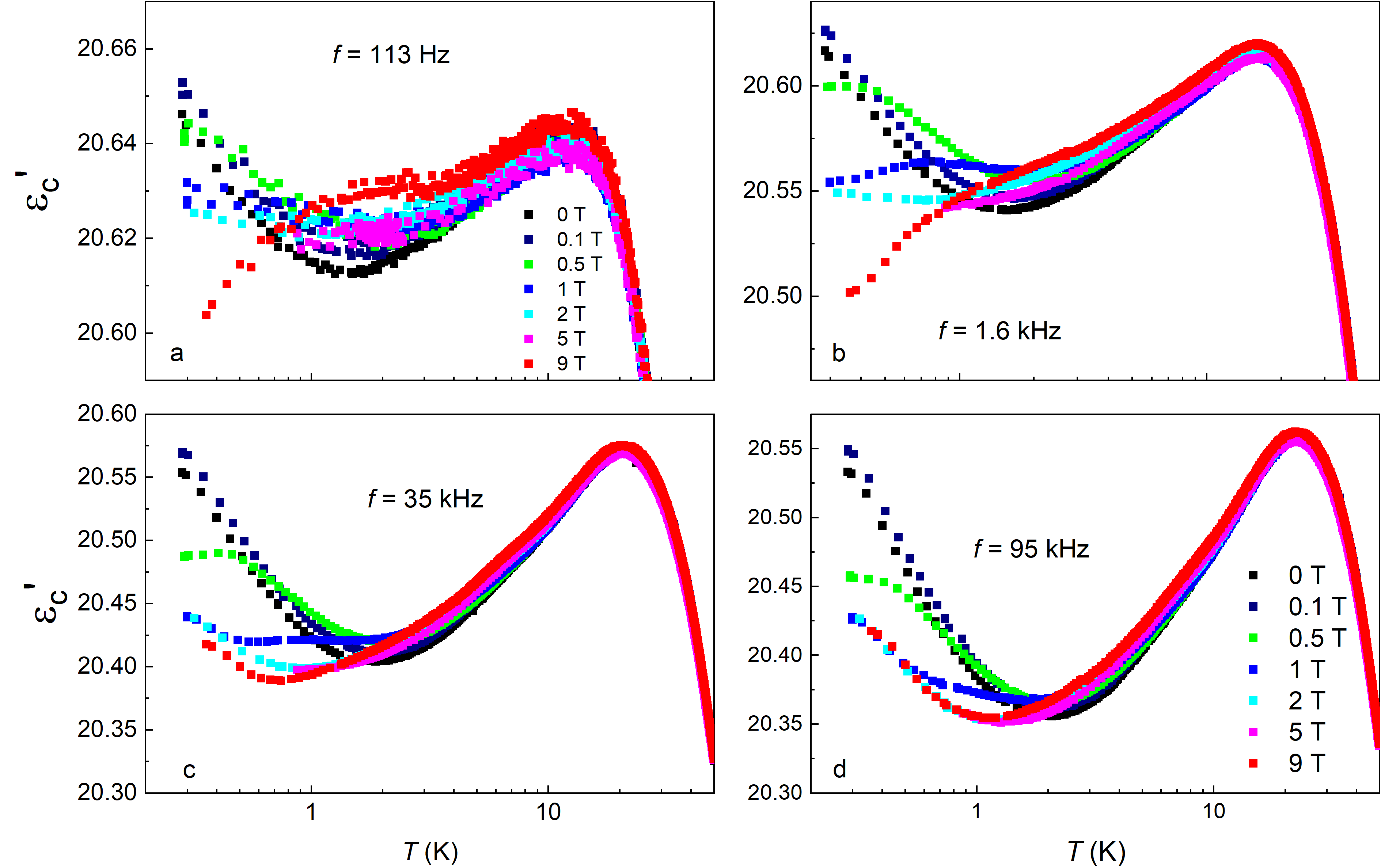}
  \caption{Temperature dependence of \(\varepsilon_c'(T)\) of \ce{NdMgAl11O19}
  in magnetic fields \(\mu_0H=0\)--9~T:
  (a) \(f=113\)~Hz, (b) \(f=1.6\)~kHz, (c) \(f=35\)~kHz, and
  (d) \(f=95\)~kHz.}
  \label{fig:Fig2_eps_vs_T_fields}
\end{figure*}

A low-temperature broad maximum in $\varepsilon_c'(T)$ may arise either from
coupling of a soft optical phonon to the acoustic
branch~\cite{Rowley2014ferroelectric}, as recently established in
SrTiO$_3$~\cite{orenstein2025SrTiO3}, or from coupling to
defects~\cite{vendik1997modeling,fujishita2016quantum}. In the present material,
the relatively modest value of \(\varepsilon_c'\), compared with classical
large-permittivity quantum paraelectrics, does not favor a strong
optical--acoustic phonon coupling scenario. In the related magnetoplumbite
compound \ce{EuAl12O19}, the relevant polar phonon remains above
\(\sim 3\)~THz~\cite{Bastien2024}, which also makes such a scenario less likely
here.

Below about 30~K, however, the data show a reproducible deviation from the pure
Barrett form. In view of the previously established \ce{Nd^{3+}} CEF scheme, we
interpret this feature as the dielectric signature of thermal depopulation of
the first excited CEF doublet. We therefore fit the permittivity using a sum of
the Barrett background and an additional term associated with this
excited-doublet contribution [see Appendix~\ref{app:derivation}],
\begin{equation}
\varepsilon_c'(T)=\varepsilon'_{\rm Barrett}(T)+\Delta\varepsilon'_{\rm CEF}(T),
\label{eq:eps_total}
\end{equation}
with
\begin{equation}
\Delta\varepsilon'_{\rm CEF}(T)=\frac{A}{T}\,\sech^2\!\left(\frac{\Delta}{2T}\right),
\label{eq:cef_depop}
\end{equation}
where \(\Delta\) represents the energy separation of the first excited doublet
expressed in kelvin units, while \(A\) sets the magnitude of its dielectric
contribution projected onto the measured axis, with
\(\Delta\varepsilon'_{\rm CEF}(T)\to 0\) as \(T\to 0\).

A fit of the zero-field $\varepsilon_c'(T)$ in Fig.~\ref{fig:Fig1_eps}(a)
using Eqs.~(\ref{eq:barrett})--(\ref{eq:cef_depop}) over the full measured
range \(0.3\)--\(285\)~K yields \(\varepsilon_{\infty}=18.44 \pm 0.01\),
\(T_{1}=147.90 \pm 0.90\)~K, \(T_{0}=5.23 \pm 1.18\)~K,
\(C=140.87 \pm 2.07\)~K, \(A=2.80 \pm 0.05\)~K, and
\(\Delta=25.85 \pm 0.32\)~K. Reducing the upper fitting limit to lower
temperatures shifts the Barrett parameters somewhat---in particular, \(T_1\)
decreases by a few kelvin and \(T_0\) becomes negative---but the fitted value
of \(\Delta\) remains robust against the choice of fitting window. The
resulting gap, \(\Delta=25.85 \pm 0.32\)~K, agrees well with the lowest
\ce{Nd^{3+}} crystal-electric-field excitation previously estimated from
point-charge calculations, \(E_{1}\simeq 2.17\)~meV
(\(\approx 25.2\)~K)~\cite{Kumar2025_NdMgAl}. This agreement supports direct
assignment of the 25--30~K dielectric anomaly to thermal depopulation of the
first excited CEF doublet. The dielectric response therefore suggests three
superposed components: (i)~a Barrett-like host background, (ii)~an additional
largely frequency-independent contribution on the 25--30~K scale associated
with the first excited CEF doublet, and (iii)~a low-temperature dispersive
channel discussed below.

The frequency dependence of the real permittivity is illustrated in
Fig.~\ref{fig:Fig1_eps}(b) and Fig.~\ref{fig:Fig1_eps}(c) up to 50~K in
\(\mu_0H=0\) and 9~T, respectively. At elevated temperatures, the permittivity
is nearly frequency independent. Below \(\sim 30\)--35~K, however, the response
develops a clear frequency dependence. This dispersion is consistent with the
type of disorder-related mechanism discussed previously for \ce{CeMgAl11O19},
namely local inhomogeneity associated with random Mg/Al site
mixing~\cite{Kumar2025CeMgAl11O19ME}. This dispersive contribution should be
distinguished from the deviation of the zero-field temperature dependence from
the pure Barrett law, which is captured here by the additional term associated
with thermal depopulation of the first excited CEF doublet. The weak frequency
dependence of the 25--30~K anomaly position, together with the stability of the
fitted gap against the fitting window, supports treating this feature separately
from the broader dispersive response that develops below \(\sim 30\)--35~K. As
in \ce{EuAl12O19}, one may also consider a contribution from correlated antipolar
dipolar dynamics~\cite{Bastien2024}; however, in the present case the agreement
between the fitted \(\Delta\) and the previously inferred first excited CEF level
supports assignment of the 25--30~K feature to the first excited-state CEF
doublet, while the accompanying dispersion is consistent with disorder broadening
of the response.

In addition, the zero-field temperature dependence in
Fig.~\ref{fig:Fig1_eps}(a) shows a further upturn of \(\varepsilon_c'(T)\)
below about 2~K, indicating another low-temperature contribution beyond both
the broader dispersion observed below \(\sim 30\)~K and the anomaly on the
25--30~K scale. A qualitatively similar low-temperature upturn has also been
observed in \ce{CeMgAl11O19} and \ce{PrMgAl11O19}, suggesting that this feature
may reflect a broader aspect of the magnetoplumbite hexaaluminate family rather
than a phenomenon unique to the Nd
compound~\cite{Kumar2025CeMgAl11O19ME,Kumar2026PrMgAl11O19ME}. The onset of
this upturn near 2~K is comparable to the energy scale set by the Curie--Weiss
temperature \(\Theta_{\rm CW} \approx -0.5\)~K~\cite{Kumar2025_NdMgAl},
suggesting that the development of antiferromagnetic correlations among the
\ce{Nd^{3+}} moments plays a role. In this family, rare-earth ions are known to
exhibit local off-centering within the oxygen bipyramidal
site~\cite{Kumar2025,Kumar2025_NdMgAl,Kumar2026SmMgAl11O19,Bastien2025}, which
modifies the Nd--O--Nd superexchange geometry and generates a distribution of
exchange interactions. Within this picture, the sub-2~K dielectric upturn and
its associated dispersion arise from antiferromagnetically correlated
\ce{Nd^{3+}} clusters whose slow internal dynamics---governed by the distributed
exchange landscape---couple to the polarizable host lattice and produce a
frequency-dependent enhancement of \(\varepsilon_c'\).

The field evolution of the dielectric response is shown in
Fig.~\ref{fig:Fig2_eps_vs_T_fields}, where \(\varepsilon_c'(T)\) is plotted for
\(\mu_0H=0\)--9~T at representative frequencies. The broad feature near
\(\sim 25\)--30~K remains visible throughout the measured field range. By
contrast, the low-temperature response becomes increasingly field and frequency
dependent. Below about 2~K, \(\varepsilon_c'(T)\) varies substantially across
the measured frequency window, with the differences becoming more pronounced at
higher fields. For fields up to \(\mu_0H=0.5\)~T, the low-temperature
permittivity shows the same qualitative trend at all measured frequencies, namely
an upturn upon cooling. For fields of \(\mu_0H \geq 1\)~T, this behavior becomes
distinctly frequency dependent. In particular, at \SI{9}{\tesla} the
low-temperature upturn is suppressed at lower frequencies
[Figs.~\ref{fig:Fig2_eps_vs_T_fields}(a) and
\ref{fig:Fig2_eps_vs_T_fields}(b)], whereas it reappears at higher frequencies
[Figs.~\ref{fig:Fig2_eps_vs_T_fields}(c) and
\ref{fig:Fig2_eps_vs_T_fields}(d)].

\begin{figure*}
  \centering
  \includegraphics[width=1.0\textwidth]{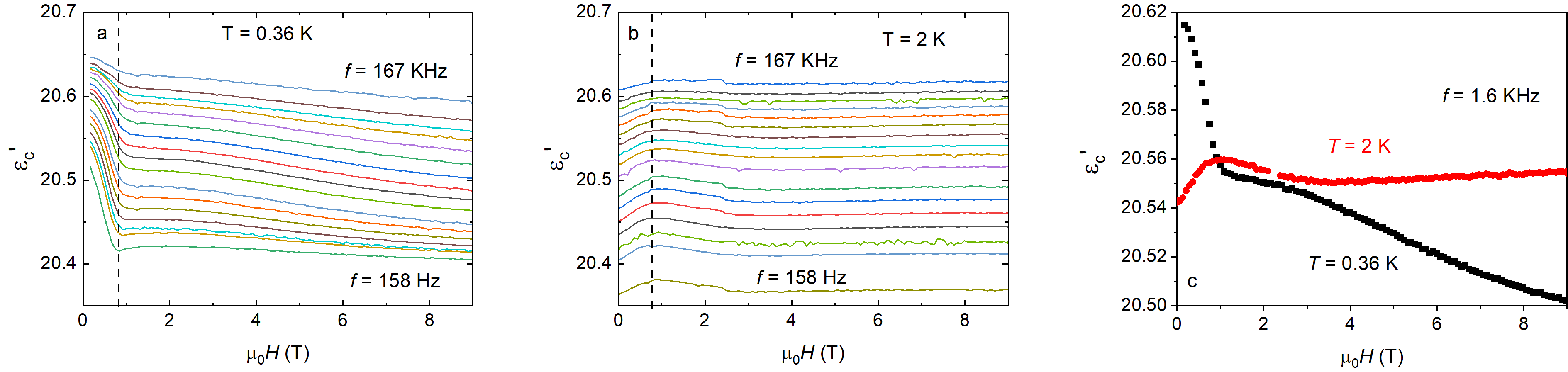}
  \caption{Field dependence of the real permittivity \(\varepsilon_{c}'(H)\) of
  \ce{NdMgAl11O19}. Panels (a) and (b) show measurements at several frequencies
  between 158~Hz and 167~kHz at (a) \(T=0.36\)~K and (b) \(T=2\)~K. Panel (c)
  compares \(\varepsilon_{c}'(H)\) measured at the selected frequency
  \(f=1.6\)~kHz for \(T=0.36\)~K and \(T=2\)~K. The vertical dashed line marks
  the crossover field \(\mu_0H_c \simeq 0.85\)~T.}
  \label{fig:Fig3_eps_vs_H}
\end{figure*}

Further information is obtained from isothermal field sweeps.
Figure~\ref{fig:Fig3_eps_vs_H} shows \(\varepsilon_c'(H)\) at
\(T\simeq 0.36\)~K and 2~K for several frequencies between 158~Hz and 167~kHz.
At both temperatures, the data show a change in field dependence near
\(\mu_0H_c \simeq 0.85\)~T across the full measured frequency window. At 0.36~K
this low-field scale appears as a pronounced knee separating a steep initial
suppression of \(\varepsilon_c'\) from a much weaker high-field evolution. At
2~K the response over the same field range becomes broader and changes sign, with
an initial low-field enhancement followed by flattening at higher field. In
addition to the primary crossover near \(\mu_0H_c\), a weaker deviation from the
smooth high-field trend is visible at both temperatures just below \(\sim 3\)~T,
appearing as a small kink or shoulder in the field dependence.

\section{Discussion}
\label{sec:discussion}

The results above reveal two characteristic low-energy scales superposed on a
Barrett-like host background. We now discuss their interpretation, beginning
with the 25--30~K anomaly and then turning to the low-temperature field-dependent
response.

The close agreement between the fitted gap \(\Delta = 25.85 \pm 0.32\)~K and
the first excited \ce{Nd^{3+}} CEF doublet at \(E_1 \simeq 2.17\)~meV
(\(\approx 25.2\)~K) from point-charge analysis~\cite{Kumar2025_NdMgAl} strongly
supports assignment of the 25--30~K dielectric anomaly to thermal depopulation
of this doublet. Such CEF--dielectric coupling is consistent with the broader
observation that rare-earth CEF excitations can leave clear signatures in optical
and magnetodielectric response through coupling to phonons and other low-energy
modes~\cite{PhysRevLett.122.177601,PhysRevResearch.2.043169,%
PhysRevResearch.2.023162}. Since the Zeeman energy accessible up to 9~T remains
well below this CEF scale, the weak field dependence of the 25--30~K feature is
expected.

The low-temperature dielectric response involves a competition among three energy
scales whose relative importance changes with temperature:
(i)~antiferromagnetic exchange correlations among the \ce{Nd^{3+}} moments,
enhanced by the distribution of exchange paths arising from rare-earth
off-centering; (ii)~Zeeman splitting of the ground-state Kramers doublet; and
(iii)~the thermal energy \(k_BT\).

At \(T = 0.36\)~K, the system lies well below \(|\Theta_{\rm CW}| \approx
0.5\)~K (\(\Theta_{\rm CW} \approx -0.5\)~K~\cite{Kumar2025_NdMgAl}), placing
it within the regime of frustrated antiferromagnetic correlations on the
triangular \ce{Nd^{3+}} sublattice. In this regime, thermal broadening alone
cannot account for the observed crossover: the Cole--Cole broadening parameter
\(\alpha\) extracted from the zero-field permittivity remains nearly temperature
independent (\(\alpha \simeq 0.60\)--0.65 between 0.3 and 10~K; see
Appendix~\ref{app:colecole}), whereas a static disorder distribution of
activation energies would yield \(\alpha \to 1\) as
\(T \to 0\)~\cite{Phillips1987}. A weakly temperature-dependent \(\alpha\) of
comparable magnitude is instead characteristic of broadly distributed cluster or
glassy dynamics in correlated magnets~\cite{Phillips1987,Bastien2024}. The
pronounced suppression of \(\varepsilon_c'\) at 0.36~K under field application
up to \(\mu_0H_c\) [Fig.~\ref{fig:Fig3_eps_vs_H}(a)] then provides direct
evidence that antiferromagnetic correlations contribute to the low-field
permittivity, since static structural disorder would not be appreciably
suppressed by such modest fields. We interpret this contribution as arising from
antiferromagnetically correlated \ce{Nd^{3+}} clusters---facilitated by the
distribution of exchange interactions due to rare-earth
off-centering~\cite{Kumar2025,Kumar2025_NdMgAl,Kumar2026SmMgAl11O19,%
Bastien2025}---whose internal spin dynamics modulate the local dielectric
response through exchange-mediated coupling to the polarizable host
lattice~\cite{Bastien2024,PhysRevResearch.2.023162}.

The frequency dependence of this low-temperature contribution provides a key
constraint on its microscopic origin. A static spin--phonon coupling mechanism,
in which antiferromagnetic correlations soften an optical phonon and thereby
enhance \(\varepsilon_c'\) via the Lyddane--Sachs--Teller relation, would
produce an essentially frequency-independent shift across our measurement window,
since the relevant transverse-optic mode in the magnetoplumbite hexaaluminates
lies above \(\sim 3\)~THz~\cite{Bastien2024}---more than seven decades above our
highest measurement frequency. The strong dispersion observed below 2~K therefore
rules out a purely magnetoelastic (soft-mode) origin and instead points to slow
relaxational dynamics of the correlated-cluster manifold, whose characteristic
frequencies fall within the Hz--kHz range probed here. The Zeeman energy at the
crossover, \(\Delta E_Z = g_c\mu_{\rm B}\mu_0H_c \approx 2.1\)~K (using
\(g_c \simeq 3.7\) from Ref.~\cite{Kumar2025_NdMgAl}), is comparable to the
correlation energy scale expected for a frustrated triangular antiferromagnet
with \(\Theta_{\rm CW} \approx -0.5\)~K~\cite{Ramirez2025,Balents2010}, and we
accordingly identify \(\mu_0H_c\) as the field at which Zeeman splitting
overcomes the antiferromagnetic correlations and drives the system out of the
correlated regime.

By contrast, at \(T \simeq 2\)~K the isothermal magnetization follows a
Brillouin-function form for an effective spin-\(\tfrac{1}{2}\)
doublet~\cite{Kumar2025_NdMgAl}, indicating that antiferromagnetic correlations
already play a reduced role and the response is closer to a single-ion limit. At
this temperature, thermal energy enters as the third competing scale: the field
response becomes broader, the low-field response changes sign, and the crossover
near \(\mu_0H_c\) loses the sharp knee seen at 0.36~K. The smooth evolution from
sharp at 0.36~K to broad at 2~K thus reflects the progressive dilution of
correlation effects by thermal fluctuations.

In the high-field regime (\(\mu_0H \gtrsim \mu_0H_c\)), the dielectric response
is governed by the Zeeman-split ground-state Kramers doublet. Simple population
redistribution between the two field-split states cannot drive the slow,
frequency-dependent dispersion observed here, since the static two-level
susceptibility is essentially frequency independent over the Hz--kHz
range~\cite{Phillips1987}. Instead, the field dependence reflects the dynamics
of inter-state transitions---moment-flipping events---between the Zeeman-split
branches, whose characteristic rate is Boltzmann-suppressed as
\(\exp(-\Delta E_Z/k_BT)\) and therefore decreases rapidly with increasing
field. This is directly observed in the Cole--Cole analysis, where at
\(\mu_0H = 9\)~T the relaxation frequency drops by more than an order of
magnitude on cooling below 2~K while the broadening parameter decreases to
\(\alpha \approx 0.42\) (see Appendix~\ref{app:colecole}). The simultaneous
slowing and partial Debye-like sharpening of the relaxation is consistent with
a progressively better-defined single-ion moment-flipping process.

The primary crossover near \(\mu_0H_c\) does not exhaust the full high-field
response. In Fig.~\ref{fig:Fig3_eps_vs_H}(c), a weaker shoulder-like deviation
is visible at both temperatures just below \(\sim 3\)~T. A natural
interpretation is that \(\mu_0H_c \simeq 0.85\)~T marks the principal exit from
the correlated regime, whereas the feature near 2--3~T reflects a more gradual
approach to a fully Zeeman-split single-ion limit in which the field-induced
splitting exceeds the residual broadening from local CEF disorder and weak
effective exchange. These are not two distinct thermodynamic boundaries but
rather two stages of the same evolution of a broadened low-energy manifold under
field.

Structural and spectroscopic considerations support such a broadened picture.
Local structural disorder has been reported for \ce{NdMgAl11O19} and related
magnetoplumbite hexaaluminates from SCXRD studies, which reveal local deviations
from an ideal single-site rare-earth
environment~\cite{Kumar2025_NdMgAl,Kumar2025,Kumar2026SmMgAl11O19}. Rare-earth
CEF spectra are highly sensitive to such local variations: distinct local charge
surroundings can produce additional CEF
excitations~\cite{Goremychkin2011SpatialInhomogeneity}, while smooth
environmental disorder broadens the spectrum into a wide
envelope~\cite{Vayer2024EntropyStabilized}. Exchange and fluctuation effects add
further linewidth~\cite{Jensen1991RareEarthMagnetism}. In \ce{NdMgAl11O19}, the
relevant low-energy limit is therefore a ground-state doublet manifold broadened
by distributed local CEF parameters, strain fields, and exchange couplings
arising from rare-earth off-centering, on top of which the antiferromagnetic
correlations develop at the lowest temperatures.

This behavior differs from that of the isostructural Kramers compound
\ce{CeMgAl11O19}, where the dielectric anomaly appears near
\(T^{*}\approx 3.6\)~K and was discussed in terms of higher-order virtual CEF
processes rather than direct thermal
depopulation~\cite{Kumar2025CeMgAl11O19ME}. The contrast is sharper still for
the non-Kramers compound \ce{PrMgAl11O19}, where off-centering of the Pr ion
lifts the non-Kramers degeneracy already in zero field, producing a distribution
of low-energy scales and a broader dielectric and thermodynamic
response~\cite{Kumar2025,Kumar2026PrMgAl11O19ME}. In \ce{NdMgAl11O19}, the
ground-state doublet remains Kramers-protected in zero field, but its
experimental manifestation is broadened by disorder and, at the lowest
temperatures, by antiferromagnetic correlations. Comparison with \ce{EuAl12O19},
where the dielectric response is dominated by the frustrated \ce{AlO5} dipolar
subsystem with no rare-earth CEF control~\cite{Bastien2024}, completes the
sequence: \ce{EuAl12O19} represents the dipolar host, \ce{CeMgAl11O19} shows
indirect CEF control, \ce{PrMgAl11O19} shows a distributed non-Kramers
response, and \ce{NdMgAl11O19} provides a Kramers reference system in which
both excited-doublet depopulation and a field-driven crossover from correlated to
single-ion behavior are reflected in the permittivity.

\section{Conclusion}

Dielectric spectroscopy of single-crystalline \ce{NdMgAl11O19} reveals three
superposed contributions to the permittivity along the \(c\) axis. At higher
temperatures, \(\varepsilon_c'(T)\) follows a Barrett-like quantum-paraelectric
background. Below \(\sim 30\)~K, a robust additional contribution with
\(\Delta = 25.85 \pm 0.32\)~K is assigned to thermal depopulation of the first
excited \ce{Nd^{3+}} CEF doublet~\cite{Kumar2025_NdMgAl}. At the lowest
temperatures, the response becomes strongly frequency and field dependent. The
crossover at \(\mu_0H_c \simeq 0.85\)~T reflects a competition between
antiferromagnetic correlations---whose relaxational, cluster-like character is
evidenced by the strong frequency dependence and the near-constant Cole--Cole
broadening parameter---and Zeeman splitting of the ground-state Kramers doublet.
\ce{NdMgAl11O19} thus serves as a Kramers reference system for separating
excited-state CEF signatures from those of the field-tuned,
correlation-dominated ground-state sector in centrosymmetric frustrated
rare-earth hexaaluminates.

\section*{Acknowledgments}
We acknowledge funding from Charles University in Prague within the Primus
research program (Grant No.~PRIMUS/22/SCI/016), the Grant Agency of Charles
University (Grant No.~438425), the Czech Science Foundation (Project
No.~24/10791S) and the project TERAFIT -
CZ.02.01.01/00/22\_008/0004594 co-financed by European Union and the Czech
Ministry of Education, Youth and Sports. Crystal growth, structural analysis,
and magnetic property measurements were carried out at MGML, supported within
the Czech Research Infrastructures program (Project No.~LM2023065).

\appendix

\section{Dielectric contribution from the first excited CEF doublet}
\label{app:derivation}

In the main text, the deviation of $\varepsilon_c'(T)$ from a pure Barrett
background below $\sim 30$\,K is assigned to the dielectric response associated
with thermal depopulation of the first excited \ce{Nd^{3+}} CEF doublet at
$E_1 \simeq 2.17$\,meV ($\approx 25.2$\,K)~\cite{Kumar2025_NdMgAl}. To
parameterize this contribution, we use the form
\begin{equation}
\Delta\varepsilon'_{c,\mathrm{CEF}}(T)
  = \frac{A_c}{T}\,\sech^2\!\left(\frac{\Delta}{2T}\right),
\label{eq:cef_depop_si}
\end{equation}
which follows from a minimal two-level approximation for the relevant low-energy
CEF sector, as derived below.

We consider two CEF states separated by an energy $\Delta$, with an effective
dipolar coupling to the electric field \(\vec{E}\parallel c\). Following
standard canonical-ensemble statistical
mechanics~\cite{Reif2009,LandauLifshitz2013}, a minimal Hamiltonian is
\begin{equation}
\mathcal{H}(E) = \frac{\Delta}{2}\sigma_z - p E\,\sigma_z,
\end{equation}
with eigenenergies \(E_\pm(E) = \pm(\Delta/2 - pE)\), where $p$ denotes an
effective dipole component along the measured axis. The partition function is
\begin{equation}
Z(E) = 2\cosh\!\left[\beta\!\left(\tfrac{\Delta}{2}-pE\right)\right],
\end{equation}
with $\beta = (k_B T)^{-1}$. The linear dielectric susceptibility follows from
the second field derivative of the free-energy
density~\cite{Phillips1987},
\begin{equation}
\chi_{c,\mathrm{CEF}}
  = \frac{n\,p^2}{k_B T}\,\sech^2\!\left(\frac{\Delta}{2k_BT}\right),
\end{equation}
where $n = N/V$ is the number density of the relevant units. The corresponding
contribution to the relative permittivity is
\begin{equation}
\Delta\varepsilon'_{c,\mathrm{CEF}}(T)
  = \frac{n\,p^2}{\varepsilon_0 k_B T}
    \,\sech^2\!\left(\frac{\Delta}{2k_BT}\right),
\end{equation}
which yields Eq.~(\ref{eq:cef_depop_si}) upon expressing $\Delta$ and $T$ in
kelvin units (absorbing $k_B$), with $A_c = np^2/(\varepsilon_0 k_B)$.

This expression captures the leading temperature dependence expected from
thermal depopulation of a discrete excited doublet. The off-diagonal tunneling
term is omitted because for a Kramers doublet the diagonal Zeeman coupling
dominates; at zero field the form captures the purely thermal population effect.
Equation~(\ref{eq:cef_depop_si}) gives Curie-like behavior ($\Delta\varepsilon'
\propto 1/T$) for $T\gg\Delta$, and activated suppression
($\propto 4e^{-\Delta/T}$) for $T\ll\Delta$. The fitted value
$\Delta = 25.85 \pm 0.32$~K is consistent with $E_1 \simeq 2.17$~meV
($\approx 25.2$~K)~\cite{Kumar2025_NdMgAl}.

\section{Dielectric loss and Cole--Cole analysis}
\label{app:colecole}

\begin{figure*}
 \centering
 \includegraphics[width=0.48\textwidth]{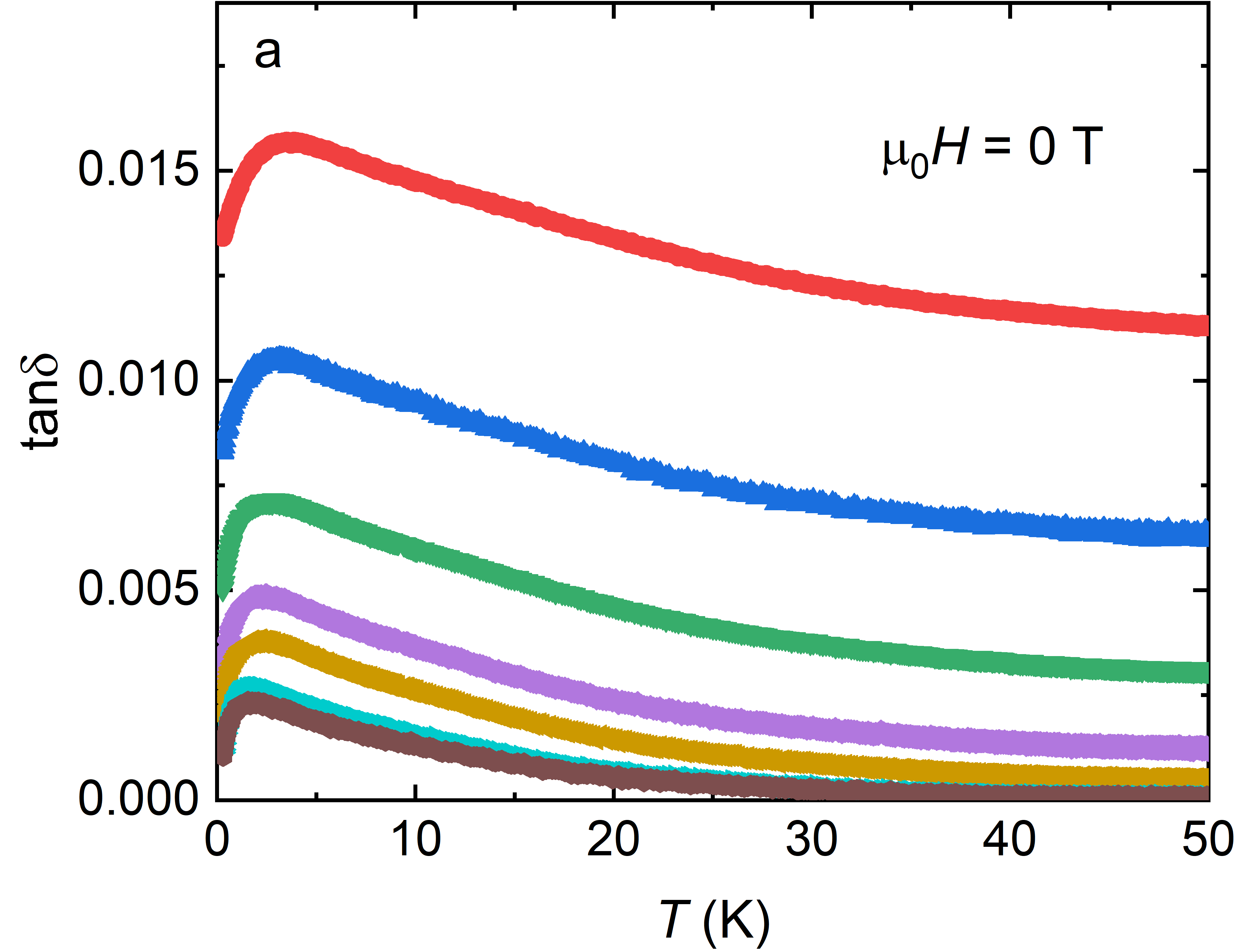}\hfill
 \includegraphics[width=0.48\textwidth]{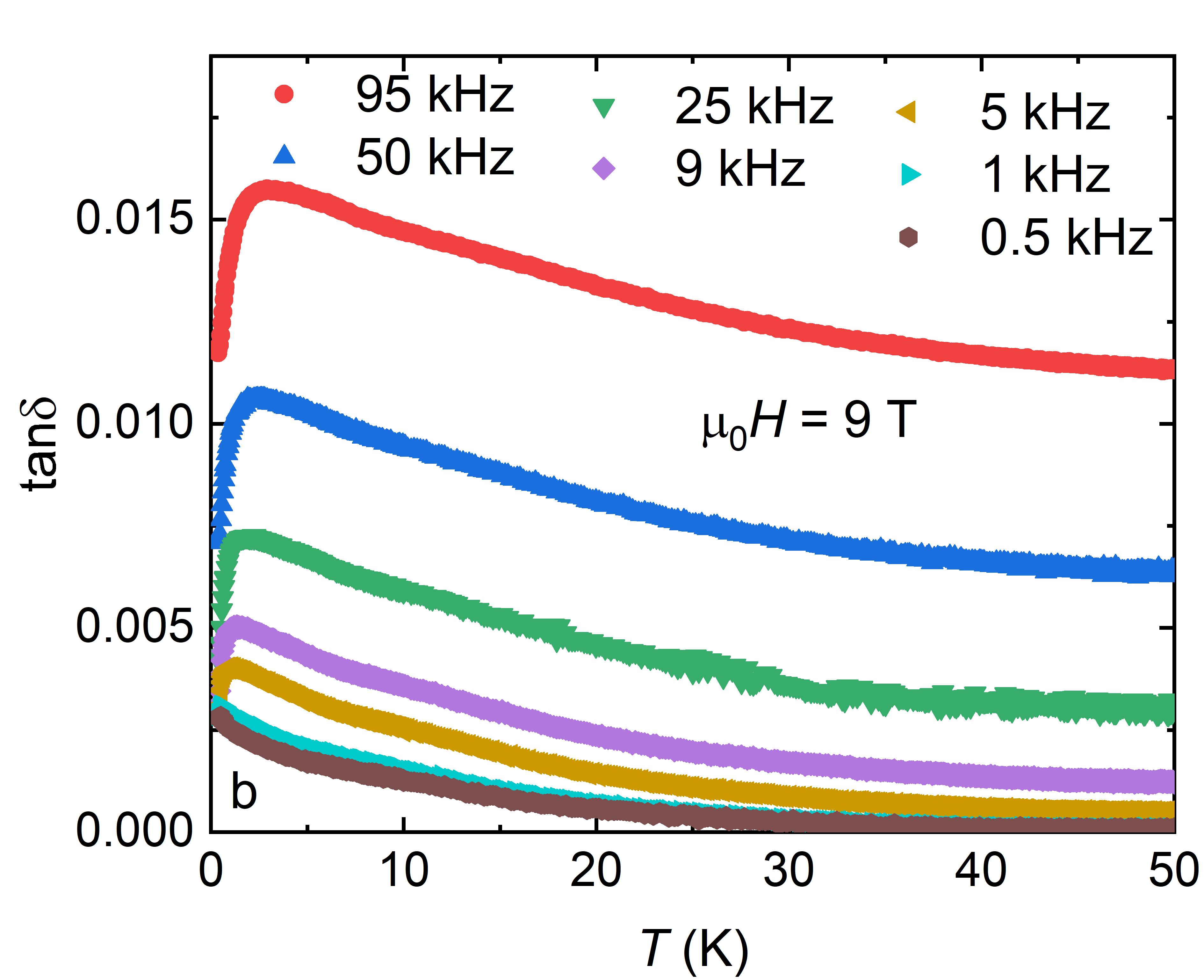}
 \caption{Temperature dependence of the loss tangent \(\tan\delta(T)\) of
 \ce{NdMgAl11O19} measured at various frequencies with
 \(E\parallel H\parallel c\) in (a)~\(\mu_0H=0\)~T and (b)~\(\mu_0H=9\)~T.}
 \label{fig:tandelta_0T_9T}
\end{figure*}

Figure~\ref{fig:tandelta_0T_9T} shows the temperature dependence of the loss
tangent \(\tan\delta(T)\) measured in zero field and in \(\mu_0H=9\)~T with
\(H\parallel c\). Dielectric losses increase monotonically with increasing
frequency, indicating that the relaxation frequency lies above our highest
frequency of 50~kHz. The losses decrease at temperatures below 2\,K, but this
decrease is lost in a magnetic field at the lowest frequencies, qualitatively
indicating a decrease in the relaxation frequency with increasing magnetic field.

We analyzed the real part of the permittivity \(\varepsilon'_c(f)\) using the
Cole--Cole formula
\begin{equation}
\varepsilon_c^\ast(\omega)=\varepsilon_{c,\infty}+
\frac{\Delta\varepsilon_c}{1+(i\omega\tau)^{1-\alpha}},
\label{eq:colecole_complex_si}
\end{equation}
where \(\omega=2\pi f\), \(\varepsilon_{c,\infty}\) is the high-frequency
permittivity limit, \(\Delta\varepsilon_c\) is the dielectric strength, \(\tau\)
is the characteristic relaxation time, and \(\alpha\) is the Cole--Cole
broadening parameter ($\alpha=0$: ideal Debye; $0 < \alpha \leq 1$: distributed
relaxation times). Spectra were fitted for all temperatures between 0.3 and 10~K
and all applied fields.

\begin{figure*}[t]
\centering
\includegraphics[width=0.48\textwidth]{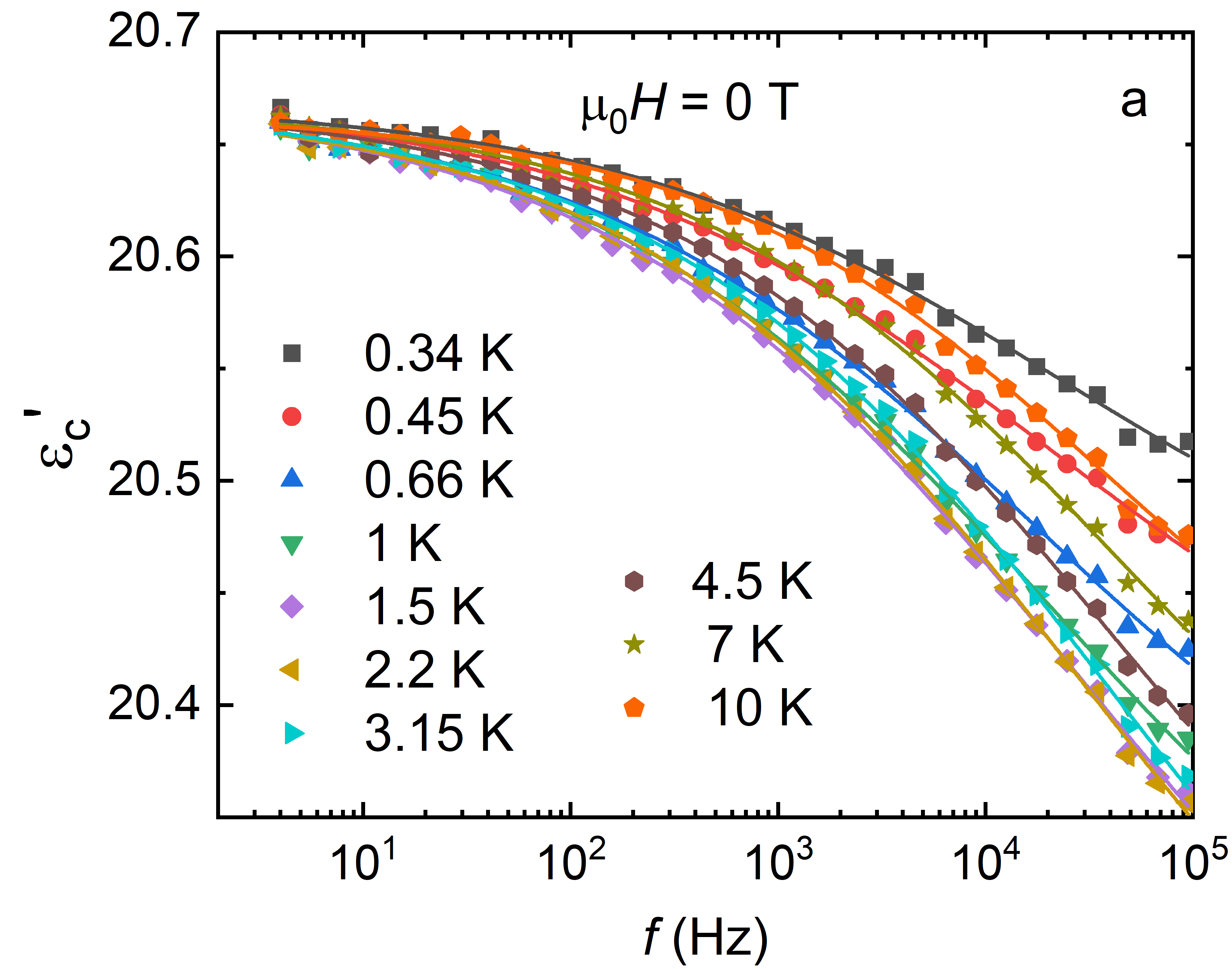}\hfill
\includegraphics[width=0.48\textwidth]{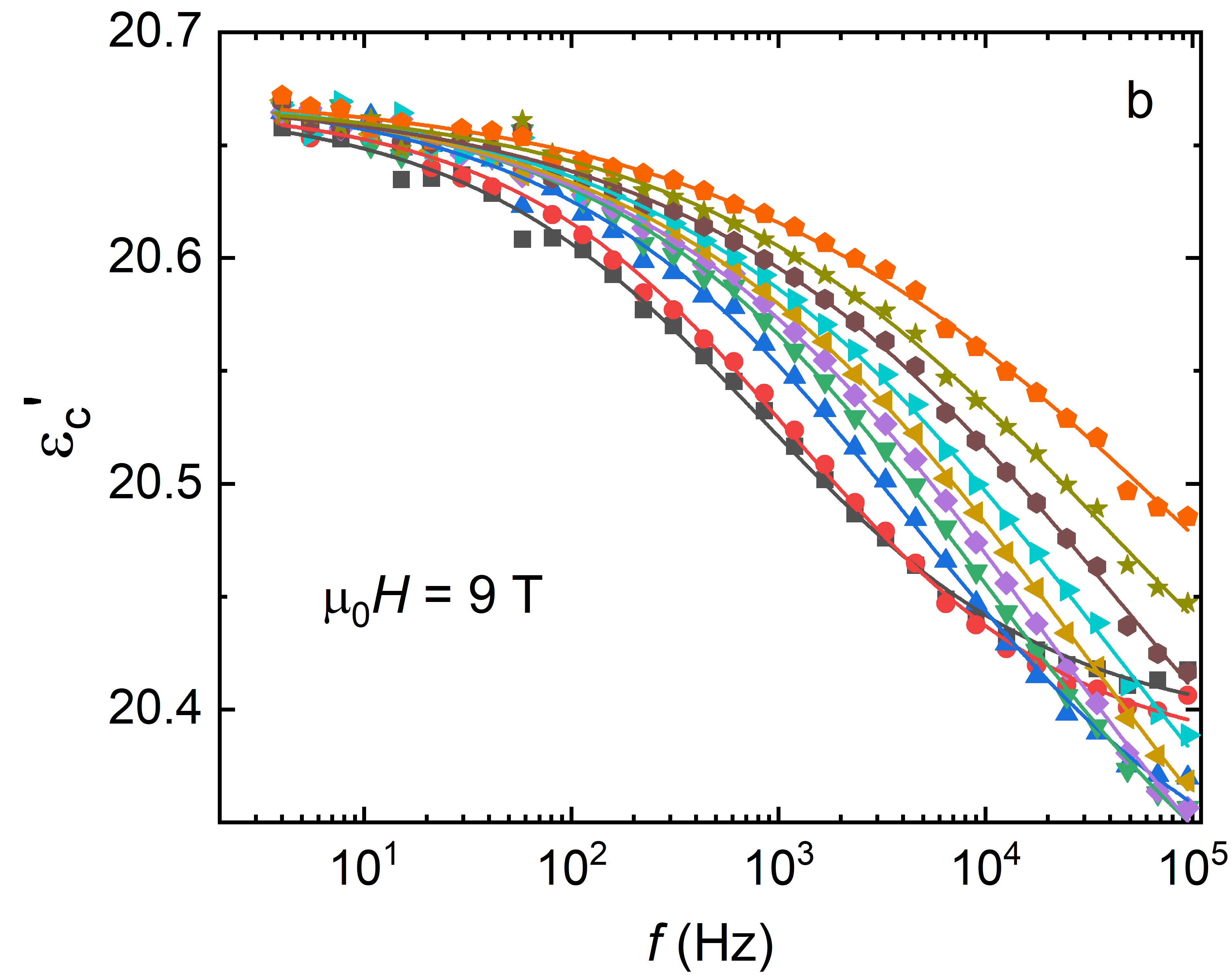}\\[0.6em]
\includegraphics[width=0.48\textwidth]{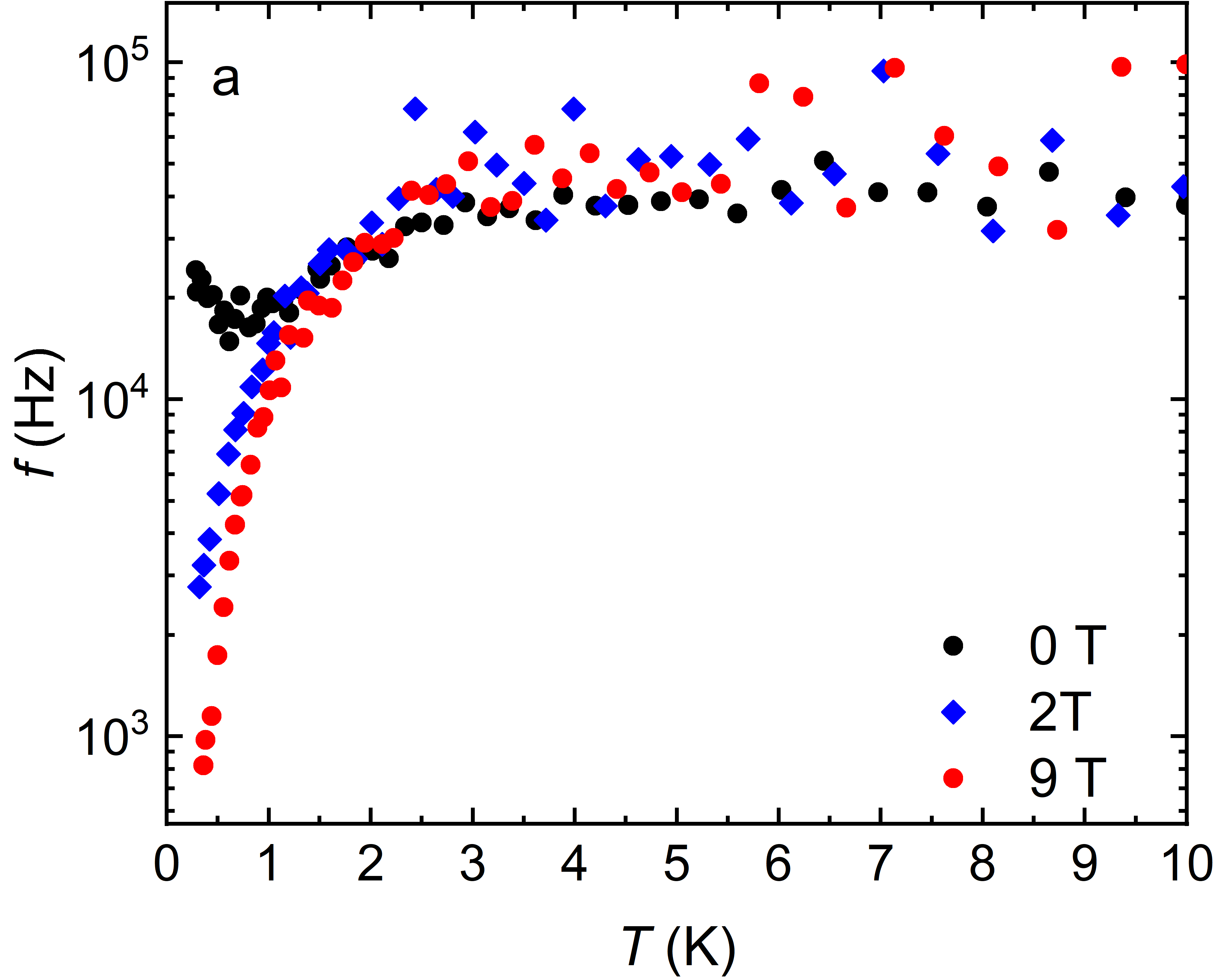}\hfill
\includegraphics[width=0.48\textwidth]{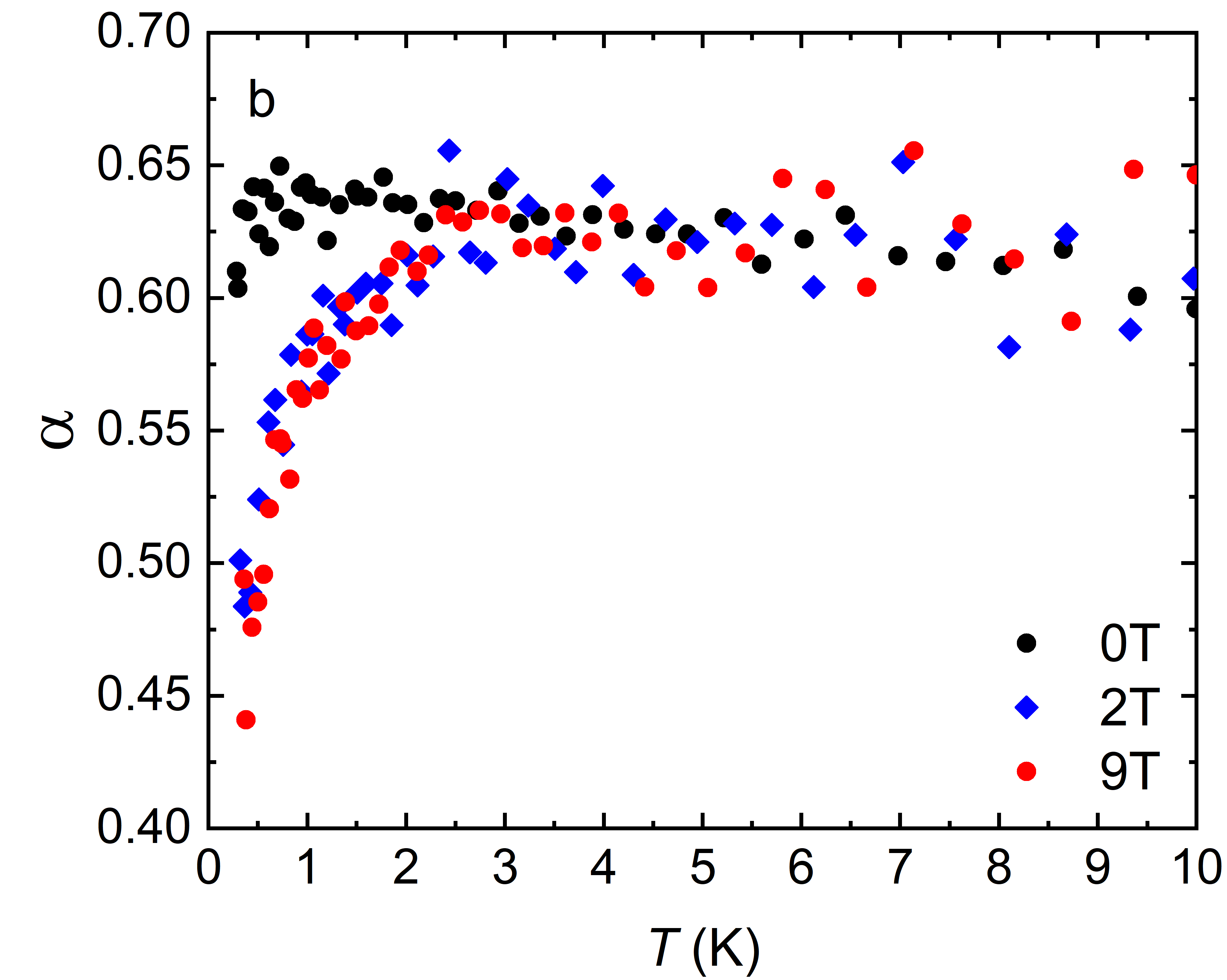}
\caption{Cole--Cole analysis of the frequency-dependent real permittivity of
\ce{NdMgAl11O19}. Panels~(a) and~(b) show representative fits of
\(\varepsilon'_c(f)\) using the Cole--Cole form
[Eq.~(\ref{eq:colecole_complex_si})] at selected temperatures in \(\mu_0H=0\)
and 9~T, respectively. Panels~(c) and~(d) show the temperature dependence of the
characteristic relaxation frequency \(f_{\mathrm{rel}}=1/(2\pi\tau)\) and the
Cole--Cole broadening parameter \(\alpha\), respectively, extracted from fits in
\(\mu_0H=0\), 2, and 9~T.}
\label{fig:colecole_fits}
\end{figure*}

The Cole--Cole analysis reveals a clear magnetic-field dependence of the
dielectric relaxation. In fields above \(\sim 0.5\)~T, the characteristic
relaxation frequency \(f_{\mathrm{rel}}=1/(2\pi\tau)\) decreases markedly on
cooling below \(\sim 2\)~K, indicating a pronounced slowing down of the
dielectric dynamics in the Zeeman-split single-ion regime. At the same time, the
fitted broadening parameter \(\alpha\) decreases, reaching values close to
\(\alpha\approx 0.42\) at \(\mu_0H=9\)~T and \(T=0.3\)~K. In contrast, for
fields below 1~T, \(\alpha\) stays in the range \(\sim 0.60\)--0.65 over the
whole temperature interval 0.3--10~K and \(f_{\mathrm{rel}}\) decreases below
3\,K only slightly.

In the low-field regime, the relaxation remains strongly non-Debye and broadly
distributed, compatible with an unresolved ground-state Kramers doublet in the
presence of disorder and antiferromagnetic correlations. The weak temperature
dependence of \(\alpha\) in zero field is itself significant: for a purely static
distribution of activation energies, the spread in \(\ln\tau\) space scales as
$\Delta\varepsilon/(k_BT)$ and would drive \(\alpha \to 1\) as
\(T \to 0\)~\cite{Phillips1987}. The observed near-constancy of \(\alpha
\simeq 0.60\)--0.65 down to 0.3~K is instead characteristic of broadly
distributed cluster or glassy dynamics in correlated magnets, in which the
distribution of relaxation times arises not from static disorder alone but from
antiferromagnetic correlations~\cite{Phillips1987}. This is consistent with the
lowest measured temperatures lying below \(|\Theta_{\rm CW}| \approx
0.5\)~K~\cite{Kumar2025_NdMgAl}.

At higher magnetic field, the Zeeman splitting of the originally doubly
degenerate level resolves the two branches. The relaxation becomes slower but
less broadly distributed as \(\alpha\) decreases, consistent with a
progressively better-defined single-ion moment-flipping process: transitions
between the two field-split states become Boltzmann-suppressed (\(\propto
\exp(-\Delta E_Z/k_BT)\)), slowing \(f_{\mathrm{rel}}\), while the distribution
of relaxation times narrows as the system approaches a uniform single-ion limit.
This crossover is consistent with the same field range where the magnetic
specific heat evolves toward a Schottky-like anomaly associated with the
Zeeman-split ground-state doublet~\cite{Kumar2025_NdMgAl}.

The Cole--Cole parameter also provides information about the effective energy
scale governing the low-temperature relaxation dynamics. For a thermally
activated process with \(\ln\tau \propto \varepsilon/k_BT\), the half-width of
the relaxation-time distribution in \(\ln\tau\) space maps onto an effective
dynamical energy scale \(W \approx \pi\alpha k_BT/(1-\alpha)\)~\cite{Phillips1987}.
The near-constancy of \(\alpha \simeq 0.62\) implies that \(W\) scales linearly
with temperature, meaning that the energy landscape explored by the relaxing
degrees of freedom tracks the available thermal energy---as expected for
correlated dynamics within a dense manifold of nearly degenerate states on a
frustrated lattice, rather than relaxation over a fixed static energy barrier. At
\(T = 0.36\)~K, this gives \(W/k_B \approx 1.9\)~K, characterizing the
effective energy scale of the antiferromagnetically correlated cluster dynamics
at this temperature. The Zeeman splitting at the crossover field, \(\Delta E_Z =
g_c\mu_{\rm B}\mu_0 H_c \approx 2.1\)~K (using \(g_c \simeq 3.7\) from
Ref.~\cite{Kumar2025_NdMgAl}), exceeds this dynamical scale, confirming that
\(\mu_0H_c \simeq 0.85\)~T marks the field at which Zeeman splitting freezes out
the correlated relaxation and drives the system toward a field-split single-ion
regime.

\section{Schematic of the broadened low-energy CEF sector}
\label{app:schematic}

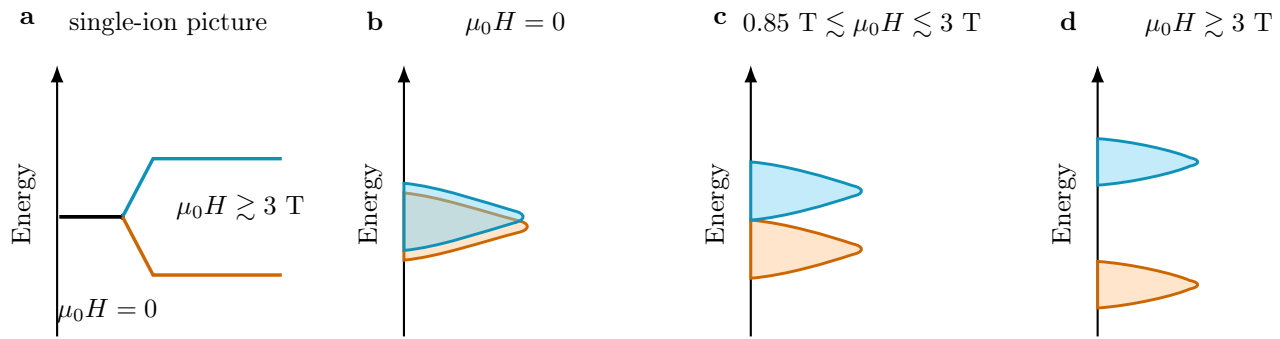
\begin{figure*}[t]
\centering
\begin{tikzpicture}[x=0.58cm,y=0.86cm,>=Latex]

\def\pw{7.4}
\def\ph{5.8}
\def\gapx{0.55}

\coordinate (A) at (0,0);
\coordinate (B) at ({\pw+\gapx},0);
\coordinate (C) at ({2*(\pw+\gapx)},0);
\coordinate (D) at ({3*(\pw+\gapx)},0);

\tikzset{
  panellabel/.style={font=\bfseries\fontsize{10}{12}\selectfont,inner sep=3pt},
  paneltitle/.style={font=\fontsize{10}{12}\selectfont,inner sep=3pt},
  axislabel/.style={font=\fontsize{10}{12}\selectfont,inner sep=3pt},
  annolabel/.style={font=\fontsize{10}{12}\selectfont,inner sep=3pt}
}

\newcommand{\DOSLobe}[6]{%
\filldraw[fill=#5,fill opacity=0.45,draw=#6,line width=1.1pt]
($(#1)+(1.15,{#2+#3})$)
.. controls ($(#1)+(1.95,{#2+0.90*#3})$) and ($(#1)+(2.95,{#2+0.45*#3})$) ..
($(#1)+({1.15+0.90*#4},{#2+0.18*#3})$)
.. controls ($(#1)+({1.15+1.00*#4},{#2+0.10*#3})$) and
            ($(#1)+({1.15+1.00*#4},{#2-0.10*#3})$) ..
($(#1)+({1.15+0.90*#4},{#2-0.18*#3})$)
.. controls ($(#1)+(2.95,{#2-0.45*#3})$) and ($(#1)+(1.95,{#2-0.90*#3})$) ..
($(#1)+(1.15,{#2-#3})$) -- cycle;
}

\newcommand{\PanelTitle}[3]{%
\node[anchor=north west,panellabel] at ($(#1)+(0.10,\ph-0.10)$) {#2};
\node[anchor=north,paneltitle] at ($(#1)+(0.5*\pw,\ph-0.10)$) {#3};
}

\PanelTitle{A}{a}{single-ion picture}
\draw[->,thick] ($(A)+(1.15,0.55)$) -- ($(A)+(1.15,4.75)$);
\node[rotate=90,axislabel] at ($(A)+(0.38,2.60)$) {Energy};
\draw[line width=1.4pt] ($(A)+(1.20,2.40)$) -- ($(A)+(2.65,2.40)$);
\fill ($(A)+(2.65,2.40)$) circle (0.05);
\draw[cyan!70!black,line width=1.3pt]
  ($(A)+(2.65,2.40)$) -- ($(A)+(3.35,3.30)$) -- ($(A)+(6.30,3.30)$);
\draw[orange!80!black,line width=1.3pt]
  ($(A)+(2.65,2.40)$) -- ($(A)+(3.35,1.50)$) -- ($(A)+(6.30,1.50)$);
\node[annolabel] at ($(A)+(2.3,0.95)$) {$\mu_0H=0$};
\node[annolabel] at ($(A)+(5.35,2.55)$) {$\mu_0H\gtrsim 3~\mathrm{T}$};

\PanelTitle{B}{b}{$\mu_0H=0$}
\draw[->,thick] ($(B)+(1.15,0.55)$) -- ($(B)+(1.15,4.75)$);
\node[rotate=90,axislabel] at ($(B)+(0.38,2.60)$) {Energy};
\DOSLobe{B}{2.25}{0.52}{2.90}{orange!45}{orange!80!black}
\DOSLobe{B}{2.40}{0.52}{2.80}{cyan!45}{cyan!70!black}

\PanelTitle{C}{c}{$0.85~\mathrm{T}\lesssim \mu_0H \lesssim 3~\mathrm{T}$}
\draw[->,thick] ($(C)+(1.15,0.55)$) -- ($(C)+(1.15,4.75)$);
\node[rotate=90,axislabel] at ($(C)+(0.38,2.60)$) {Energy};
\DOSLobe{C}{1.90}{0.45}{2.60}{orange!45}{orange!80!black}
\DOSLobe{C}{2.80}{0.45}{2.60}{cyan!45}{cyan!70!black}

\PanelTitle{D}{d}{$\mu_0H\gtrsim 3~\mathrm{T}$}
\draw[->,thick] ($(D)+(1.15,0.55)$) -- ($(D)+(1.15,4.75)$);
\node[rotate=90,axislabel] at ($(D)+(0.38,2.60)$) {Energy};
\DOSLobe{D}{1.35}{0.36}{2.35}{orange!45}{orange!80!black}
\DOSLobe{D}{3.25}{0.36}{2.35}{cyan!45}{cyan!70!black}

\end{tikzpicture}
\caption{Qualitative schematic of broadening and field evolution in the
low-energy CEF sector of \ce{NdMgAl11O19}. Panel~(a) shows an idealized
single-ion picture without broadening: a degenerate zero-field ground-state
doublet evolves into two Zeeman-split branches with increasing field.
Panels~(b)--(d) illustrate the corresponding broadened low-energy response in
the presence of disorder and weak interactions. In panel~(b), the zero-field
response appears as an unresolved broadened doublet. In panel~(c), the applied
field separates the low-energy manifold into two partially overlapping broadened
branches. In panel~(d), the same two branches are more clearly resolved at
higher field, approaching a more distinct single-ion limit. The schematic is
qualitative and is not intended as a microscopic fit to the data.}
\label{fig:schematic_doublet_evolution}
\end{figure*}

To illustrate the interpretation adopted in the main text,
Fig.~\ref{fig:schematic_doublet_evolution} shows a qualitative schematic of how
the low-energy \ce{Nd^{3+}} CEF sector may appear in the presence of local
structural disorder and weak effective interactions. This picture is motivated by
crystallographic disorder reported for the magnetoplumbite hexaaluminate
family~\cite{Kumar2025_NdMgAl,Kumar2025,Gao2026CeMgAl11O19,Kumar2026SmMgAl11O19,%
Tu2024PrMgAl11O19,Cao2024PrMgAl11O19}.

At zero field, the ground-state Kramers doublet remains symmetry protected, but
local structural variations and weak low-energy correlations can broaden its
experimental manifestation into a distribution of nearby local realizations
[Fig.~\ref{fig:schematic_doublet_evolution}(b)]. With increasing field, the
Zeeman energy progressively separates the two components. For intermediate
fields, the splitting is comparable to the broadening scale and the two branches
still overlap substantially [Fig.~\ref{fig:schematic_doublet_evolution}(c)]. At
higher field the splitting clearly exceeds the broadening
[Fig.~\ref{fig:schematic_doublet_evolution}(d)] and the response approaches a
more distinct single-ion limit.

\bibliography{main}

\end{document}